# Examining Rail Transportation Route of Crude Oil in the US Using Crowdsourced Social Media Data[1]

**Yuandong Liu, Ph.D.**
[1]Department of Transportation Engineering
Chang'an University, Xi'an, Shanxi, China, 710064
[2]Transportation Analytics and Decision Sciences Group
Oak Ridge National Laboratory, Oak Ridge, Tennessee, USA, 37830
ORCiD: 0000-0001-8404-5844
Email: yuandong_liu@hotmail.com

**Majbah Uddin, Ph.D.**
Transportation Analytics and Decision Sciences Group
Oak Ridge National Laboratory, Oak Ridge, Tennessee, USA, 37830
ORCiD: 0000-0001-9925-3881
Email: uddinm@ornl.gov

**Shih-Miao Chin, Ph.D.**
Transportation Analytics and Decision Sciences Group
Oak Ridge National Laboratory, Oak Ridge, Tennessee, USA, 37830
Email: chinc@ornl.gov

**Ho-Ling Hwang, Ph.D.**
Transportation Analytics and Decision Sciences Group
Oak Ridge National Laboratory, Oak Ridge, Tennessee, USA, 37830
Email: hwanghl@ornl.gov

**Jiaoli Chen, Ph.D.**
Transportation Analytics and Decision Sciences Group
Oak Ridge National Laboratory, Oak Ridge, Tennessee, USA, 37830
Email: jiaolichen@gmail.com

*Revised Manuscript Submitted [March 10, 2023]*

**Data Accessibility Statements:** The data that support the findings of this study are openly available in figshare at https://doi.org/10.6084/m9.figshare.22227124.v1, reference number 26. The authors confirm that the data supporting the findings of this study are available within the article or its supplementary materials.

---

[1] This manuscript has been authored by UT-Battelle, LLC, under contract DE-AC05-00OR22725 with the US Department of Energy (DOE). The US government retains and the publisher, by accepting the article for publication, acknowledges that the US government retains a nonexclusive, paid-up, irrevocable, worldwide license to publish or reproduce the published form of this manuscript, or allow others to do so, for US government purposes. DOE will provide public access to these results of federally sponsored research in accordance with the DOE Public Access Plan (http://energy.gov/downloads/doe-public-access-plan).

# ABSTRACT

Safety issues associated with transporting crude oil by rail have been a concern since the boom of US domestic shale oil production in 2012. During the last decade, over 300 crude oil by rail incidents have occurred in the US. Some of them have caused adverse consequences including fire and hazardous materials leakage. However, only limited information on the routes of crude-on-rail and their associated risks is available to the public. To this end, this study proposes an unconventional way to reconstruct the crude-on-rail routes using geotagged photos harvested from the Flickr website. The proposed method links the geotagged photos of crude oil trains posted online with national railway networks to identify potential railway segments those crude oil trains were traveling on. A shortest path-based method was then applied to infer the complete crude-on-rail routes, by utilizing the confirmed railway segments as well as their movement direction information. Validation of the inferred routes was performed using a public map and official crude oil incident data. Results suggested that the inferred routes based on geotagged photos have high coverage, with approximately 96% of the documented crude oil incidents aligned with the reconstructed crude-on-rail network. The inferred crude oil train routes were found to pass through many metropolitan areas with dense populations, who are exposed to potential risk. This finding could improve situation awareness for policymakers and transportation planners. In addition, with the inferred routes, this study establishes a good foundation for future crude oil train risk analysis along the rail route.

# INTRODUCTION

Typically, petroleum crude is transported by pipelines, which is the most cost-effective and safest mode of operation. The US domestic shale oil boom in 2012, however, left oil and gas extraction and related industries no sufficient opportunities to plan, design, build and operate much-needed pipeline infrastructures. Pipelines serving the shale oil fields, especially the Bakken shale, did not have enough capacity to carry all shale oil. Shippers, therefore, utilized trucks and railroads to transport the oil that cannot be served via pipeline. The boom of crude oil by rail transportation started in 2012 and reached its peak in 2014 (*1*). While pipelines (e.g., Dakota Access Pipeline) have been built in later years to address the demand, rail continues to serve as the second largest crude oil transportation mode in North Dakota, carrying 14% of crude oil produced in the state in 2017 according to the latest version of Freight Analysis Framework dataset (*2*). The crude oil moved by rail from North Dakota traverses a long way to petroleum refineries located on the West Coast, East Coast, as well as Gulf Coast. In addition to North Dakota, the railway also plays a role in delivering crude oil produced in Wyoming, Colorado, and several other states to the refineries.

One of the biggest concerns with crude-on-rail transportation is safety. In 2013, a train carrying crude oil derailed and exploded in Lac-Mégantic, Quebec, killing 47 people. The publicities generated by this fiery crude-on-rail mishap led to public awareness of the danger and risks associated with crude oil trains. Frequent minor incidents, however, are more difficult to get public attention than fiery accidents. From 2013 to 2021, the US has seen over 300 crude oil rail-related accidents involving several types of direct consequences, including spillage, fire, explosion, water contamination, gas dispersion, and environmental damage (*3*). Fourteen of these accidents were derailments that lead to fire, crude oil leakage, and/or people evacuation. For example, the latest crude oil train derailment that occurred in Custer, Washington in 2020, caused a fire in rail trains that burned for 8 hours. Nearly 29,000 gallons of crude oil were spilled during this incident. Aware of the possible catastrophic consequence, government agencies, researchers, and the public are eager to obtain information on the crude-on-rail operation to evaluate hazardous circumstances.

Conversely, railroad companies are conservative about their train operation information because oil train routes are considered as business sensitive. Given this situation, in 2014, the US Department of Transportation (DOT) insured an Emergency Order to safeguard crude oil shipments. That Emergency Order required railroad operating trains to submit their crude-on-rail movements, volumes, and frequencies to the State Emergency Response Commissions, when they contain "more than 1,000,000 gallons of Bakken crude oil, or approximately 35 tank cars, in a particular state" (*4*). Such information help government agencies to manage crude oil movement through their jurisdiction and evaluate the risks. Due to a confidential agreement, however, this information is not publicly available. In 2015, Wall Street Journal (WSJ) collected crude on rail movements, volumes, and frequencies information submitted by oil companies to the State Emergency Response Commissions and made it available to the concerned public (*5*). There are several limitations of the map published by WSJ, however. First, it only covered rail routes originating from Bakken shale. Second, the route information obtained is at the county level, which is not sufficient for accurate safety analysis. Moreover, due to the addition of new oil fields and changes in market demands, rail routes are subjected to changes over time. The WSJ-published map was compiled in 2015 and has not been updated since then. Thus, there is still a desire for up-to-date crude-on-rail routes at the network-link level for risk analysis needs.

To approximate unknown routes and obtain knowledge concerning crude-on-rail safety, this paper proposes an unconventional way to obtain route information using social media data, specifically, the Flickr photo-sharing website. We also integrate information from different data sources to validate the inferred route. We anticipate our findings from this research can inform the concerned public and help policymakers in relevant decision-making.

# LITERATURE REVIEW

The relevant literature discussed below is organized under two themes: the *first* part is related to crude oil transportation by rail and the *second* part reviews route determination efforts based on Flickr data.

Given the risk, associated with the transportation of crude oil by rail, to human life and the environment, special attention was given to this topic by many researchers, especially those published in the last decades. In a 2016 paper, Liu (*6*) developed a probabilistic risk analysis model and a decision support tool for estimating in-transit risk of crude oil transportation by rail in unit trains. The risk is measured by the expected number of affected persons on rail network segments (i.e., links). It is suggested that the model and tool can be used for identifying, evaluating, comparing, and prioritizing potential risk mitigation strategies. Around the same time, Oke et al. (*7*) presented a market equilibrium model for investigating strategies to mitigate the environmental and public safety risks from crude oil transportation by rail in the US. It is suggested that an integrated policy of targeted rail network capacity, pipeline investments and lifting ban of US crude oil exports will be able to address medium-term risks of crude oil transportation by rail.

Busteed (*8*) provided a comprehensive law review of crude by rail transportation dangers in terms of casualty and environmental concerns, existing regulations, and responses by rail and oil industries. Another law review was published by Gerrald and McTiernan (*9*); however, in the case of crude oil transport by rail in New York State.

Furthermore, Vaezi and Verma (*10*) analyzed rail network for crude oil transportation in Canada and potential impact on flows if a pipeline network becomes available. They proposed a bi-objective routing model that can capture both cost and risk minimization. In particular, cost objective represents total railcar-distance traveled and risk objective represents the product of population density in the vicinity of the rail links and the volume of crude oil traversing on the links. It is found that the identification of high-risk rail links could help in devising appropriate emergency responses if an incident occurs. Mason (*11*) found a positive association between the accumulation of minor incidents and the frequency of serious incidents for transportation of crude oil by rail as reported in his 2018 article. Additionally, a positive association between increased rail shipments of crude oil and the occurrence of minor incidents was found. Morrison et al. (*12*) developed a model for crude oil mode split and route assignment considering pipeline and rail modes. The model was tested in predicting changes in crude oil flow patterns and mode shares in response to changes in transportation network attributes and oil demands. Clay et al. (*13*) provided estimates of the air pollution, greenhouse gas (GHG), and spill and accident costs from long-distance movement of crude oil by rail and pipelines considering North Dakota as the shipment origins. It is found that air pollution and GHG costs are almost two times for rail compared to pipelines. Additionally, it is found that air pollution and GHG costs are higher compared to spill and accident costs.

More recently, Schneller et al. (*14*) conducted a case study for Saratoga County in the state of New York on public perceptions of technological risk, state responses, and policy for crude oil transportation by rail. The authors emphasized the risk to safety, property, and the environment of crude by rail for communities near railways that are not studied properly and the disastrous impact in the event derailment occurs. It was found that the public in the county had a low level of knowledge about oil shipments and emergency management plans. Smith (*15*) assessed regulatory compliance of crude by rail and its impact on community air pollution in Baltimore City, Maryland. It was found, from two study sites, that transportation of hazardous materials (including crude) was more prevalent during nighttime. The authors recommended further research on the detection and impacts of volatile organic compounds at sites close to rail lines that carry crude oil.

There have been a number of studies that explored the use of social media data such as geotagged Flickr photos to determine, as well as infer, transportation routes. However, these studies are mostly focused on passenger travel and, in many cases, applicable for tourists or by tourism industries. Specifically, Chareyron et al. (*16*) presented a methodology to mine tourist routes using Flickr photos. Both fastest path and most likely paths were considered to determine tourists' itinerary in the destination location. Kurashima et al. (*17*) proposed a method for travel route recommendation that uses Flickr photos to infer the location histories of tourists. One of the key model inputs was the location sequences of the person taking those geotagged photos. Sun et al. (*18*) built a travel recommendation system that is road-based and uses Flickr photos. The system provides users with the most popular landmarks as well as the best travel routes between the landmarks. Steiger et al. (*19*) proposed a framework for the purpose of detecting human mobility transportation hubs and inferring public transport flows based on geotagged social media data. Cai et al. (*20*) developed a framework that can mine frequent travel trajectory patterns as well as regions of interest based on Flickr photos. It is mentioned that the framework is able to extract expected major landmarks (e.g., cities and tourists' attractions). Spyrou et al. (*21*) analyzed user-generated routes within downtown city areas that are derived using a Flickr photo dataset. In a more recent study, Cai et al. (*22*) proposed itinerary recommender systems based on Flickr photos that consider mining of semantic trajectory pattern as well as sequential points of interests with temporal information. Yang et al. (*23*) developed a neural network that can provide travel route recommendations that are geared toward a sequence of point of interests as well as visitors' personal interests. The model was validated using a dataset consists of Flickr photos. In their 2021 paper, Kadar and Gede (*24*) proposed a methodology that uses network analysis with efficient clustering algorithms and Flickr photos. The purpose was to determine tourism flows in large-scale destination systems.

## ROUTE INFERENCE METHODOLOGY

Public geotagged crude oil train photos in the contiguous US were collected from Flickr to identify the location where crude-oil trains were sighted. Then, a methodology is proposed to identify the railroad segments these trains have traveled as well as to infer the missing routes connecting those identified railway segments.

### Harvesting Crude-oil-train Photos from Flickr

The crude-oil-train photos and videos hosted on Flickr website provide an opportunity to find the locations of crude oil trains and their routes. Flickr photos and videos contain a wide range of user-provided information including title, description, textual tags, geographical coordinates, and time stamp when the picture or video was taken. The first three data elements (i.e., title, description, and tags) give concise descriptions of photo contents, which allow for retrieval of photos that are related to a specific topic. Geographical coordinates provided are either automatically synchronized from cameras with GPS functions or provided by users (when clicking on the map) (*25*). Geographical coordinates could represent where the contents were located or where the photographer stood (*25*). In this study, we assumed the geographical coordinates of a crude-on-train photo represent the train's actual location at that time. By harvesting the geolocations of crude-oil-train photos taken in the contiguous US, information regarding where crude-oil trains were sighted, and which railroad segments they have traveled, could be obtained.

Using publicly available Flickr APIs, information from photos whose title, description or textual tags contain the keyword of "crude oil train," was collected. Photos that were taken between January 1, 2008, and June 1, 2022, were utilized for this study (*26*). Over 1,300 crude-oil-train photos taken in the contiguous US were retained. Their locations and time stamps were stored as points in an ArcGIS geodatabase and used in the next step.

**Figure 1** shows the total movements of crude oil by rail according to Energy Information Administration (EIA) statistics originating from each Petroleum Administration for Defense District (PADD) (*27*). The boom of the crude-on-rail movement in the year 2012 is clearly shown in this figure, although the crude-on-rail movement began a couple years earlier. The total crude oil by rail tonnage increased from 41,626 thousand barrels in 2011 to 144,240 thousand barrels in 2012, at a growth rate of 246%. The growth of crude oil on rail continued and peaked in 2014 with over 250 million barrels. In recent years from 2016 to 2021, however, the total crude oil by rail movements have decreased. Among all the PADDs, the largest portion of crude oil transportation by rail originated from PADD2 where Bakken shale plays a big role.

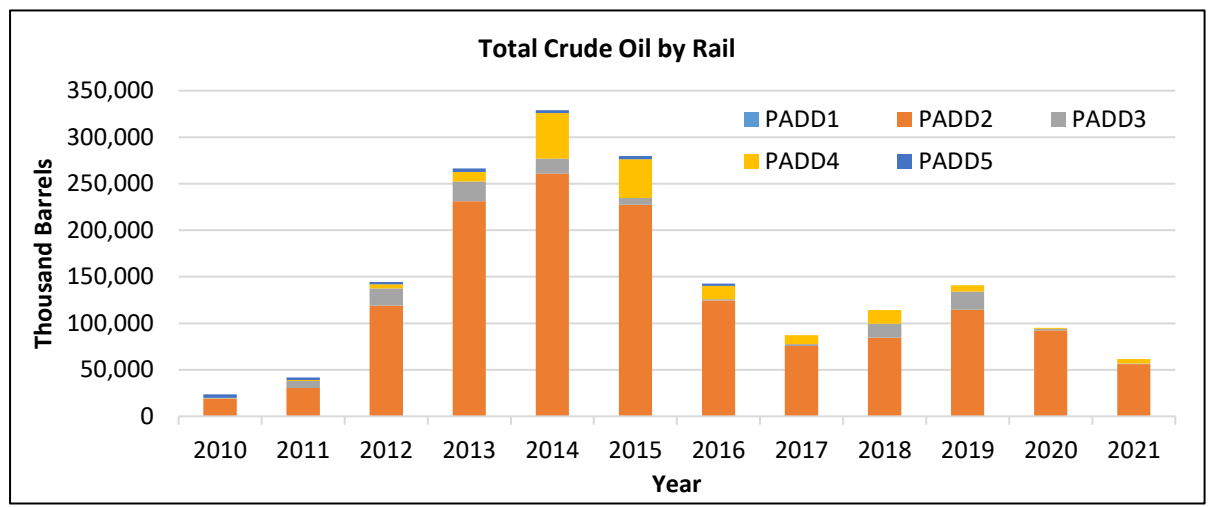

**Figure 1 Crude oil by rail received from each PADD zone over the years**

      **Figure 2** shows the temporal variation of spatial distributions of crude-on-train photos taken during the period from 2008 to 2022. Consistent with statistics reported by EIA, before 2012, there were very few crude-oil-train photos being identified. The number of photos started to increase in 2012, and clearly evident from the map, shows the Bakken oil train corridor linking North Dakota to the East Coast which dominated domestic oil train movements. Other major flows that could be observed from **Figure 2** include:
1) Bakken oil trains head south to the storage and pipeline hub in Oklahoma, as well as to the refineries in Texas and Louisiana along the Gulf Coast;
2) Bakken crude oil flow into Washington State on the West Coast, where the oil is refined or trans-loaded onto barges heading down to California; and
3) Besides Bakken oil, productions from shale deposits in Utah and Colorado can be seen from the figure.

    The temporal variations as seen in **Figure 2** might have indicated that rail movements of crude oil are relatively flexible, which allows for more frequent changes in response to new oil fields and/or market demands.

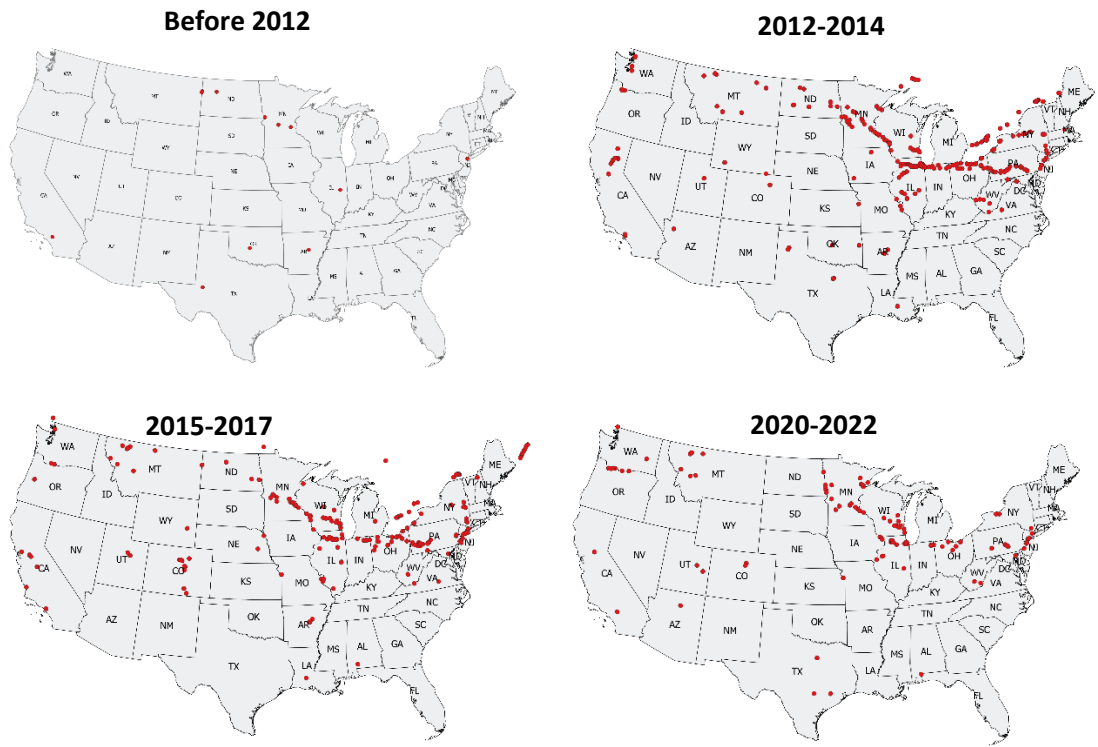

**Figure 2 Temporal variation of crude-oil-train locations derived from Flickr photos**

### Identifying Railroad Segments

By overlaying the locations of the 1300+ identified crude-oil-train photos on the North American Rail Network (NARN) *(28)*, it was found that 98% of them were located within 320 feet from their nearest railroad lines, while around 95% were within 160 feet. This suggests that the geolocations associated with crude-oil-train photos were within a close proximity to railway tracks and, therefore, are reasonably reliable sources for reflecting the locations of crude oil trains. Note that railroad companies often use main lines (rail lines excluding sidings, branches, and yards; NARN contains a data filed named NET to differentiate different types of rail lines) to deliver crude oil. This was evidenced in our research finding where 88% of crude oil trains, captured by photos, were witnessed within 160 feet of main lines, and 96% were within 320 feet. Therefore, it is reasonable and practical to use main lines to confirm and infer routes from geolocations of crude-oil-train photos.

In addition to crude-oil-train photos, rail terminals that handle loading and unloading of crude oil that are publicly available from EIA *(29)* were also evaluated. 91 crude oil rail terminals with geographical coordinates were included in the crude oil rail terminal dataset, which provides origin and destination information for crude-on-rail movements 1400 data points are identified after combing these rail terminals with Flickr photos. The route inferring algorithms were based on the 1400 data points.

The nearest railroad segment for each of the Flickr photo points/crude oil rail terminals was identified as a part of the crude-on-rail route, which is termed the "confirmed road segment." These confirmed road segments were then extended to the "confirmed route" by iteratively searching the neighboring links of the exiting confirmed road segment. The neighboring links without any branch lines at the connection point to the previously confirmed road segment will be added to the confirmed route because this is the only route the crude train can take. A simplified example is shown in **Figure 3**. First, in this example, link 4 was identified as a confirmed railroad segment in Step 1. Under Step 2, its neighboring links (link 3 and link 5) are added to the confirmed route. In Step 3, the algorithm continues to search the neighboring links of links 3 & 5. For link 3, because it connects to both link 1 & 2, the iteration stops for this direction. Link 6 is added given it connects to link 5 and does not have any branch line at the connection point. The iteration stops in this direction since link 6 is connected to the other two links. Therefore, the final confirmed route includes links 3, 4, 5 & 6.

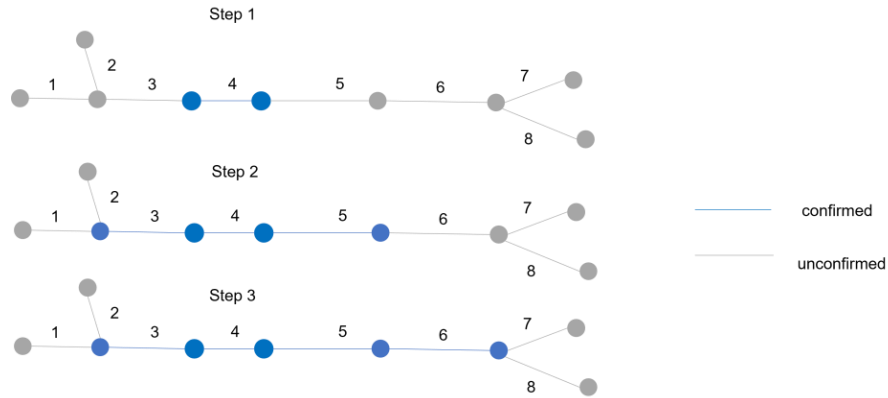

**Figure 3 Steps to extend confirmed road segment to the confirmed route**

**Figure 4** shows the results of confirmed railroad segments based on Flickr photos/rail terminals and the extended confirmed routes. As shown in this figure, after extending the confirmed road segments to confirmed routes using the proposed method, many previously disconnected road segments are now connected. For example, one crude oil train route traversing the state of Montana (MT) can be observed in **Figure 4 (b)**. This route was shown as scattered small segments in **Figure 4 (a).** However, in areas with sparse photo points, some of the confirmed routes were still not well connected, thus leaving gaps that need to be inferred.

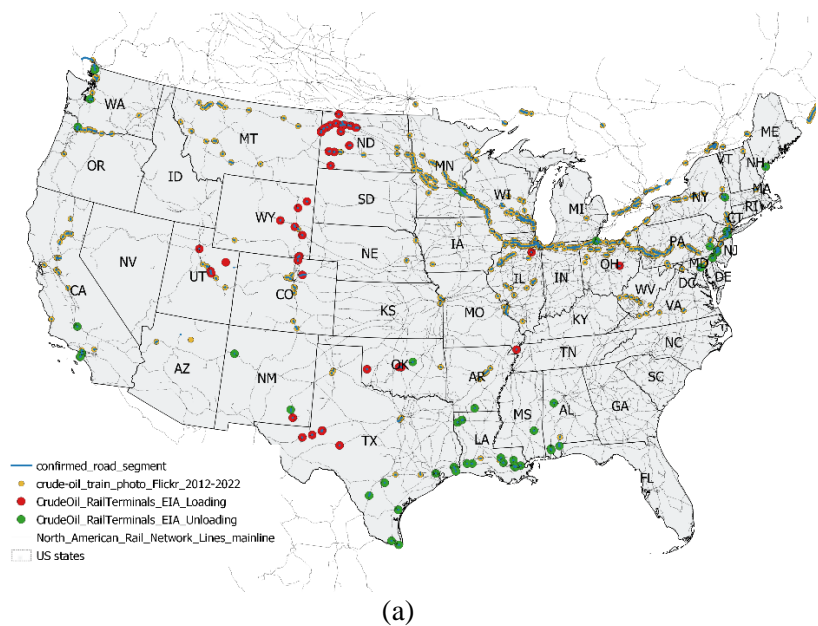

(a)

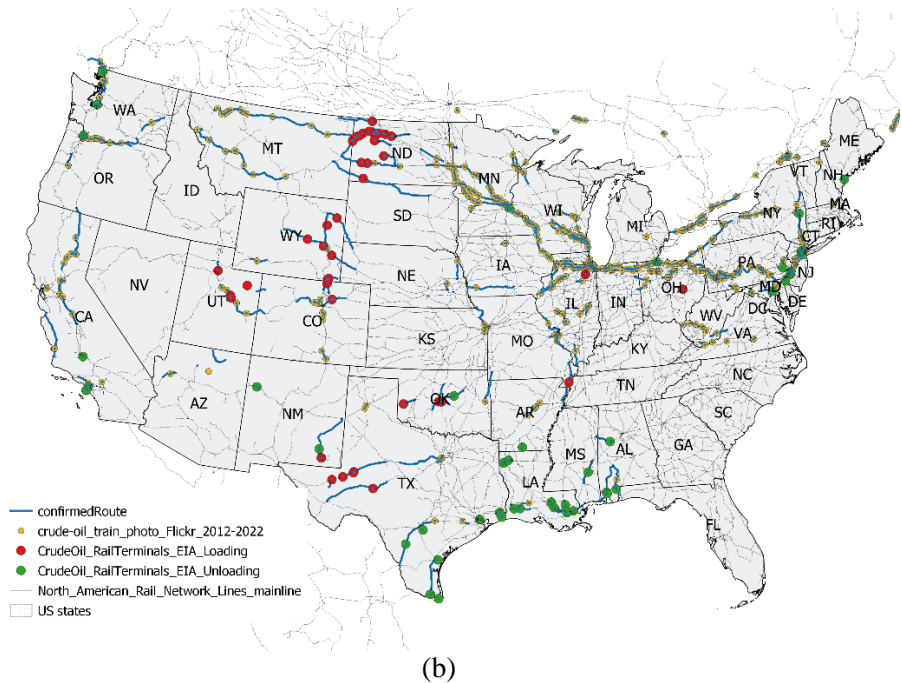

(b)

**Figure 4 (a) confirmed railroad segment; (b) confirmed route**

## Inferring Crude-on-rail Routes

EIA provides statistics and maps depicting moving directions of crude oil by rail between Petroleum Administration for Defense District (PADD) regions *(1; 27)*. **Table 1** summarize the average crude oil movements among PADDs from 2010 to 2021 *(30)*.

**TABLE 1 Annual average crude oil volumes among PADDs from 2010 to 2021(thousand barrels)**

|       | PADD1  | PADD2 | PADD3  | PADD4 | PADD5  |
|-------|--------|-------|--------|-------|--------|
| PADD1 | 0      | 0     | 0      | 0     | 0      |
| PADD2 | 48,140 | 6,846 | 24,891 | 34    | 37,283 |
| PADD3 | 11     | 1,756 | 5,822  | 678   | 997    |
| PADD4 | 1,492  | 553   | 9,841  | 45    | 1,052  |
| PADD5 | 0      | 0     | 0      | 0     | 1,805  |

**Figure 5** shows a map of the crude oil movement direction among different PADDs. The loading and unloading crude rail terminals, as well as refineries (destination of crude oil movement), are shown on the map as well.

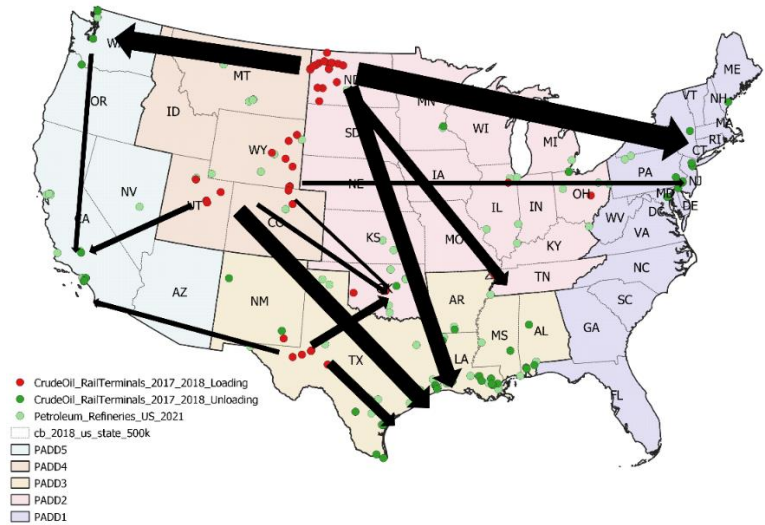

**Figure 5 Crude on rail movement directions**

Utilizing these data sources together, one can infer the missing parts of routes (identified in the section above), the following rules were applied:
(a) the inferred routes that connect the broken "confirmed routes" as well as rail terminals should be the shortest on the main line network;
(b) the inferred routes should consider the railroad ownership and trackage rights; and
(c) the inferred routes should be consistent with the direction of rail movements of crude oil as shown in **Figure 5**.

Based on the above two principles, the following methodology was designed to complete the route taken by crude oil trains:
- Step 1. Identify all broken "confirmed routes".
- Step 2. For each of these broken "confirmed routes", the shortest path algorithm is performed to identify all the possible missing routes connecting it to the other routes as well as the crude oil rail loading and unloading terminals. During identifying the shortest path between two broken "confirmed routes", the railroad ownership and trackage rights were considered. All railroad companies that either own or have trackage rights of the two confirmed routes are first identified. Then the shortest path algorithm is performed based on the railroad network that can be accessed by these companies.
- Step 3. Rank all the possible routes by distance. The one with the shortest distance is defined as the inferred route."

**Figure 6** shows the resulting crude-on-rail routes. Blue lines indicate confirmed routes and red lines indicate routes that were inferred.

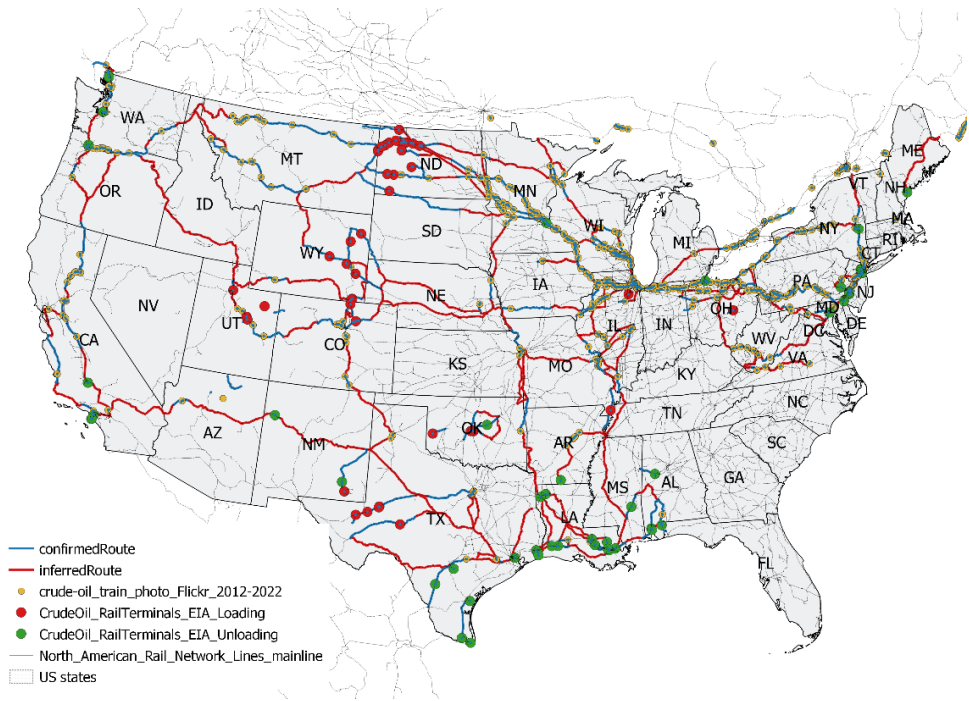

**Figure 6 Inferred routes**

## RESULTS VALIDATION AND ANALYSIS

### Results Validation
### Crude Oil Incidents

Public access to historical incident records through Pipeline and Hazardous Materials Safety Administration (PHMSA) website makes it possible to analyze the characteristics and consequences of crude-on-rail related incidents. For this research, all reported incidents associated with crude oil rail transport during the period from 2013 to 2021 were extracted from the Yearly Incident Summary Report *(3)*. **Table 2** presents an example of the crude oil incidents records. For each incident record, the city and state where the incident took place are available. Incident street address is generally not provided. Based on city and state information, geocoding is used to extract the associated coordinates of the city to represent the incident location on the map.

**TABLE 2 PHMSA crude oil incident records**

| Date | Incident Street Address | City | State | Mode of Transportation | Transportation Phase |
|---|---|---|---|---|---|
| 1/17/2013 | Unknown | Temple | TX | FRA-RAILWAY | IN TRANSIT |
| 1/25/2013 | Unknown | Kansas City | MO | FRA-RAILWAY | IN TRANSIT |
| 2/12/2013 | Unknown | BAKERSFIELD | CA | FRA-RAILWAY | STORAGE |
| 2/15/2013 | Unknown | Pine Bluff | AR | FRA-RAILWAY | IN TRANSIT |

Totally 338 incidents occurred during the period from 2013 to 2021. **Figure 7** shows the location of these incidents. Because the same coordinates were used to represent the incidents that occurred in the same city, every single point shown on the map could represent multiple incidents. Different colors (from light green to dark green) are used to indicate the incident frequency in that city. As can be seen from the figure, most of the incidents aligned well with the inferred route, except those in Arkansas, Mississippi, and Alabama. Because limited Flickr photos were taken in these three states, the inferred routes are incomplete. However, according to the frequency category, these missed incidents have low frequencies. Most of the incidents occurred only once on the missed oil route segments. In summary, among 338 incidents, less than 6% (19 incidents) were not in the surrounding areas (same city) of the inferred routes, indicating the possible coverage and accuracy of the crude oil routes reconstructed based on Flickr photos.

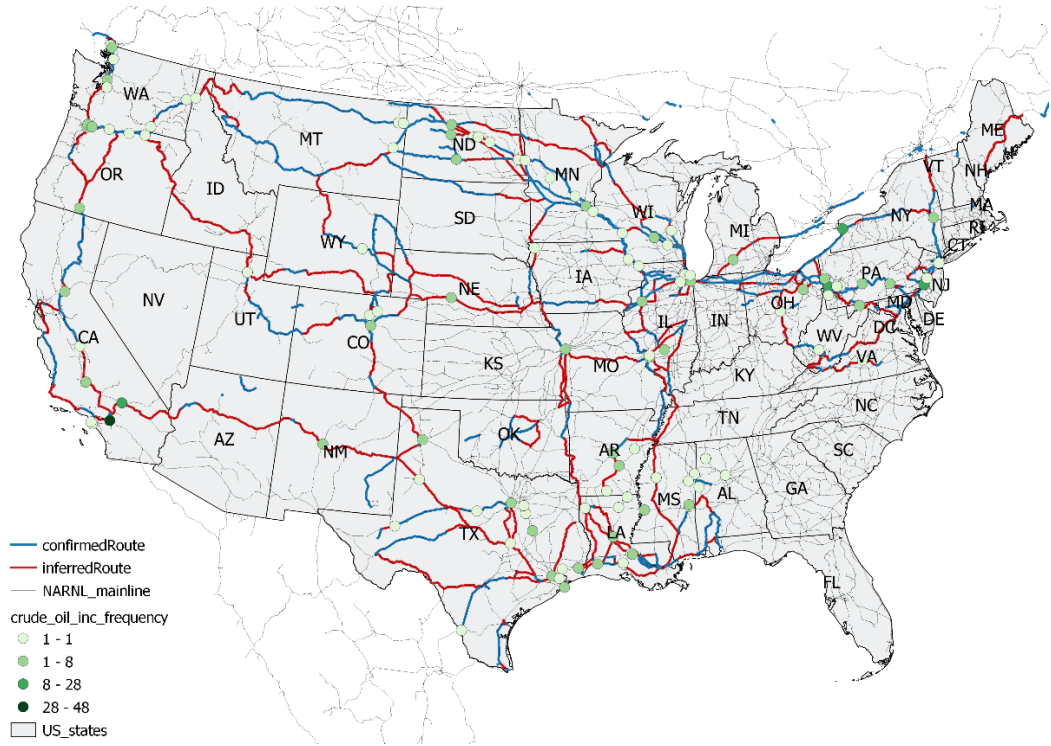

**Figure 7 Crude oil incidents on railway**

The missing 19 incidents are mainly located in the states of TX, AR, LA, MS, and AL. Because fewer Flickr photos were taken in these states, the routes were mostly inferred based on crude oil train terminals. The current route inferring algorithm connects the broken railroad segments that the crude oil train terminals and Flickr photos were on, which didn't fully utilize the origin and destination information embedded in the rail terminal dataset. In future studies, we will exploit the OD information to improve the performance of the inferring algorithm, especially in those states with fewer public photos. In addition, the crude oil accident data points were used to validate the inferred routes in this study. In the future study, they will be combined with the Flickr photos and crude oil terminal dataset to uncover more hidden routes.

The complete crude oil lines identified in this study traverse 183 metropolitan areas among all 384 metropolitan areas in the US. The crude oil trains were also found running along the Mississippi River, Susquehanna River, and Hudson River, all are important water resources to their respective neighborhoods. With the crude oil trains passing through urban areas with dense populations, a concern about population and transportation vulnerability has been raised. For instance, several Flickr photos witnessed the running of Bakken crude oil trains under a 100+ years old tunnel in Seattle downtown (*31*). The business, urban traffic, and people who work and live in that area, are highly vulnerable to the volatile crude oil flowing underground. Combining with the historical incident data, population data as well as other data sources such as crude oil movement volumes, risk analyses can be performed to assess the potential threats that crude oil trains bring to the population as well as the surrounding environment. This will be one of our future research directions.

**Public Map**

Because railroad companies are unwilling to share their oil train operation information, there are limited public data sources available. In 2015, the WSJ collected data from State Emergency Response Commissions and mapped counties traversed by trains carrying Bakken crude oil *(5)*. **Figure 8** shows an overview of the map. Color indicates crude oil train volumes – small to large as the colors range from yellow to red.

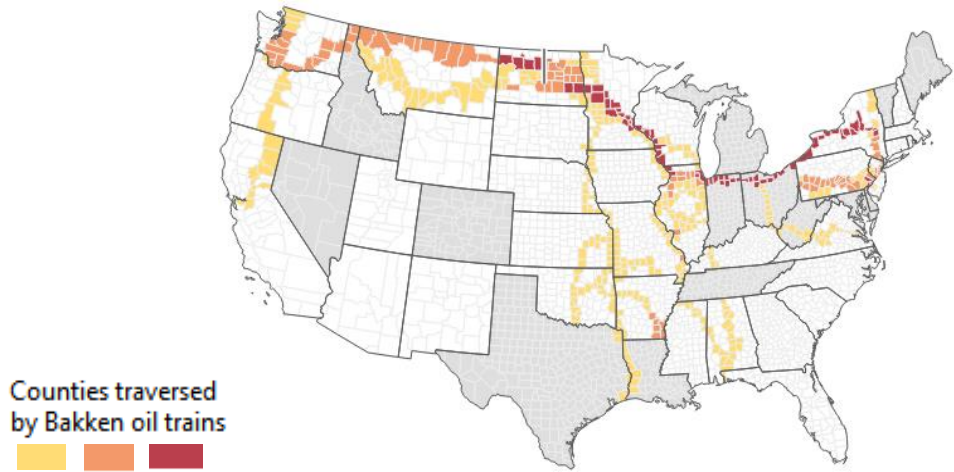

**Figure 8 county-level routes of Bakken oil trains (*5*)**

Comparing **Figure 8** to **Figure 6**, both consistency and inconsistencies were identified. Observations are summarized below:
- The complete lines of the North Dakota to East Coast corridor and North Dakota to Washington to California corridor were correctly identified. Based on **Figure 8**, these rail transportation lines have the highest Bakken oil train volumes (orange and red color).
- Another rail line from North Dakota to Texas is mostly well aligned with the inferred routes, except for the branch lines in KS, OK, MO and AR. Very few Flickr photos were taken in these four states, which limits the route inferring accuracy. However, **Figure 8** shows that the crude oil train volumes traversing these states are low (yellow). In addition, **Figure 7** shows a total of 18 accidents that occurred in these 4 states. Among all the incidents, 6 incidents that took place at the border of KS and MO, and 11 incidents that occurred within AR states were aligned with the inferred routes. Only 1 incident that occurred in AR was not on any of the inferred routes. This potentially indicates that these missing branch lines transport less crude oil volumes compared to other busy rail lines and result in fewer accidents.
- The railway lines within the state of Mississippi and Alabama were not identified, due to the same data limitation pointed out above. **Figure 7** shows that only 3 incidents occurred near the rail lines in MS and AL, indicating small crude oil train volumes traversing the rail lines.

Note that the county-level route shown in **Figure 8** is only a subset of the complete crude oil routes – they only contain those crude oil routes that originated from Bakken oil. Therefore, some inferred rail routes, such as the rail lines connecting CA and TX (**Figure 6**), were not shown in **Figure 8.**

## CONCLUSIONS

The boom of domestic shale oil production benefits US energy but also brings challenges and vulnerabilities with regard to transportation. The safety issue of moving shale oil by rail has gained much attention from the public. Knowing the routes taken by crude oil trains is critical to examining potential risks and safety concerns of the public. However, this information is typically not accessible by the public. Given this situation, our research utilized an unconventional approach to identify crude-on-rail routes with information gathered online (Flickr website). The routes were further identified with an inferring methodology developed for this study. Data sources including crude oil incidents and a public map were used to verify the inferred routes. The study results show that over 96% of documented crude oil incidents are located in the surrounding areas of the inferred route, suggesting high coverage and potential accuracy of the inferred routes. Moreover, crude-on-rail flows and routes are subjected to change with respect to new oil fields and change in market demands. With the proposed method, the route changes can potentially be tracked and identified by leveraging the latest user reports. The network developed in this study could be a good source for planners at different regional levels (e.g., city, county, and state), which could be utilized to improve their emergency management systems. The map developed could also be used to assign crude oil origin-destination volumes onto the network and estimate the crude oil movement frequency for better decision-making.

With the inferred routes, this study establishes a good foundation for many crude oil-train-related studies, such as population at-risk analysis and environmental risk assessment. It was found that crude oil trains tend to pass-through major cities with dense populations as well several major water resources. The identified routes traverse 184 metropolitan areas among 384 metropolitan areas in the US. Therefore, one of our future research directions is to analyze the population and environmental risk associated with oil trains by incorporating other data sources, such as historical crude oil incident data, population data, and crude oil movement data.

A key limitation of this study is that the crude oil-train photos on the Flickr website may not cover all the corners of crude-on-rail routes. There could be some areas where train fans hardly went to or took photos. Since no photo was captured in those areas, the inferring process applied would not have information to generate a route. Therefore, it is possible that certain routes were not covered in this study. Future research can incorporate additional public data sets to uncover those hidden routes. A few relevant public data sets are listed here:

1. **Crude oil incident data published by PHMSA** *(3)*. The incident data was used as a validation data set in this study. It can also serve as a complementary data set of the Flickr photos in future research.

2**. Flickr oil train photos.** Flickr contains higher number of photos tagged with 'oil tank' than those tagged with 'crude oil'. In this study, only train photos tagged with 'crude oil' were extracted. In future, a filtering method would be developed to extract the crude oil train photos from these 'oil train' photos to supplement the existing data sets.

3. **Railroad Picture Archives and other public figure archives.** Railroad Picture Archives (http://www.rrpicturearchives.net/default.aspx) contains over 2500 photos with 'crude oil' in its title or description, as compared to around 1300 photos in Flickr. Although the Archive might have better coverage, it provides location information at lower resolution (city level). How this dataset can be linked to the railroad network will be the main issue to be resolved before incorporating it with the existing data set.

4. **Social-media data.** Socio-media data such as Twitter (https://twitter.com) data will also be explored in the future to determine whether it contain any geo-tagged crude oil train information.

Despite this limitation, findings from this study could help policymakers in improving situation awareness and allowing better-informed decisions to be made, particularly on issues associated with those confirmed parts of crude-on-rail routes.

## AUTHOR CONTRIBUTIONS
The authors confirm their contribution to the paper as follows: study conception and design: Shih-Miao Chin, Yuandong Liu, Jiaoli Chen; data collection: Yuandong Liu, Jiaoli Chen; analysis and interpretation of results: Yuandong Liu, Jiaoli Chen, Majbah Uddin, Shih-Miao Chin; draft manuscript preparation: Yuandong Liu, Jiaoli Chen, Majbah Uddin, Ho-Ling Hwang. All authors reviewed the results and approved the final version of the manuscript.

# REFERENCES

[1] Energy Information Administration. *Crude by rail movements*. census.gov/programs-surveys/geography/technical-documentation/complete-technical-documentation/file-availability.2014.html.

[2] Federal Highway Administration. Freight Analysis Framework Version 5.In, 2021.

[3] Pipeline and Hazardous Materials Safety Administration. Yearly Incident Summary Report.In, 2021.

[4] U.S. Department of Transportation. *U.S. DOT Takes New Emergency Actions as Part of Comprehensive Strategy to Keep Crude Oil Shipments Safe*. http://www.transportation.gov/briefing-room/us-dot-takes-new-emergency-actions-part-comprehensive-strategy-keep-crude-oil.

[5] GOLD, R. Oil Trains Hide in Plain Sight.In *The Wall Street Journal*, 2015.

[6] Liu, X. Risk Analysis of Transporting Crude Oil by Rail: Methodology and Decision Support System. Transportation Research Record, 2016. 2547:57–65.

[7] Oke, O., D. Huppmann, M. Marshall, R. Poulton, S. Siddiqui. Mitigating environmental and public-safety risks of United States crude-by-rail transport. DIW Berlin Discussion Paper, 2016. Available from https://ssrn.com/abstract=2782406.

[8] Busteed, E. Bakken Crude and the Ford Pinto of Railcars: The Growing Need for Adequate Regulation of the Transportation of Crude Oil by Rail. Villanova Environmental Law Journal, 2016. 27:1–29.

[9] Gerrard, M., and E. McTiernan. Regulation of Movement of Crude oil by Rail in New York. New York Law Journal, 2015. 254:90.

[10] Vaezi, A., and M. Verma. Railroad Transportation of Crude Oil in Canada: Developing Long-Term Forecasts, and Evaluating the Impact of Proposed Pipeline Projects. Journal of Transport Geography, 2018. 69:98–111.

[11] Mason, C. Analyzing the Risk of Transporting Crude Oil by Rail. National Bureau of Economic Research, 2018. Available from http://www.nber.org/papers/w24299.

[12] Morrison, A., C. Bachmann, and F. Saccomanno. Developing an Empirical Pipeline and Rail Crude Oil Mode Split and Route Assignment Model. Transportation Research Record, 2018. 2672:261–272.

[13] Clay, K., A. Jha, N. Muller, and R. Walsh. External Costs of Transporting Petroleum Products: Evidence from Shipments of Crude Oil from North Dakota by Pipelines and Rail. The Energy Journal, 2019. 40:55–72.

[14] Schneller, A., K. Smemo, E. Mangan, C. Munisteri, C. Hobbs, and C. MacKay. Crude Oil Transportation by Rail in Saratoga County, New York: Public Perceptions of Technological Risk, State Responses, and Policy. Risks, Hazards and Crisis in Public Policy, 2020. 11:377–410.

[15] Smith, G. Assessing Regulatory Compliance and Community Air Pollution Impacts of Crude Oil by Rail Transport in Baltimore City, Maryland. Center for Advancing Research in Transportation Emissions, Energy, and Health, 2021.

[16] Chareyron, G., J. Da-Rugna, and B. Branchet. Mining Tourist Routes Using Flickr Traces. IEEE/ACM International Conference on Advances in Social Networks Analysis and Mining, 2013.

[17] Kurashima, T., T. Iwata, G. Irie, and K. Fujimura. Travel Route Recommendation using Geotagged Photos. Knowledge and Information Systems, 2013. 37: 37–60.

[18] Sun, Y., H. Fan, M. Bakillah, and A. Zipf. Road-based Travel Recommendation Using Geo-Tagged Images. Computers, Environment and Urban Systems, 2013.

[19] Steiger, E., T. Ellersiek, and A. Zipf. Explorative Public Transport Flow Analysis from Uncertain Social Media Data. ACM SIGSPATIAL International Conference on Advances in Geographic Information Systems, 2014.

[20] Cai, G., C. Hio, L. Bermingham, K. Lee, and I. Lee. Mining Frequent Trajectory Patterns and Regions-of-Interest from Flickr Photos. Hawaii International Conference on System Science, 2014.

[21] Spyrou, E., I. Sofianos, and P. Mylonas. Mining Tourist Routes from Flickr Photos.

[22] Cai, G., K. Lee, and I. Lee. Itinerary Recommender System with Semantic Trajectory Pattern Mining from Geo-Tagged Photos. Expert Systems With Applications, 2018. 94:32-40.

[23] Yang, C., J. Zhang, X. Gao, and G. Chen. MatTrip: Multi-functional Attention-based Neural Network for Semantic Travel Route Recommendation. IEEE International Conference on Web Services, 2021.

[24] Kadar, B., and M. Gede. Tourism Flows in Large-scale Destination Systems. Annals of Tourism Research, 2021. 103113

[25] Hollenstein, L., and R. Purves. Exploring place through user-generated content: Using Flickr tags to describe city cores. *Journal of Spatial Information Science*, No. 1, 2010, pp. 21-48.

[26] Liu, Y. CrudeOilTrainLoc_Flickr_2008-2022. figshare. Dataset. 2023. https://doi.org/10.6084/m9.figshare.22227340.v1

[27] Energy Information Administration. Movement of Crude Oil by Rail.In, 2021.

[28] Bureau of Transportation Statistics. North American Rail Network Lines.In, 2021.

[29] Energy Information Administration. Crude Oil Rail Terminals.In, EIA, 2019.

[30] ---. *Crude by rail movements*. census.gov/programs-surveys/geography/technical-documentation/complete-technical-documentation/file-availability.2014.html.

[31] Washington State Department of Ecology. *BNSF Bakken Crude Derailment Seattle*. https://www.flickr.com/photos/ecologywa/albums/72157645850216546.